\documentclass[aps,prd,amsmath,amssymb,superscriptaddress,twocolumn,showpacs,floatfix]{revtex4}
\usepackage{epsfig,graphics,color}
\usepackage{graphicx}
\usepackage{dcolumn}
\usepackage{bm}
\usepackage{amsmath}

\newcommand \tie {{\it i.e.}}

\newcommand \kd  {\delta}
\newcommand \ra  {\rightarrow}

\newcommand \mal {{\mathcal L}}

\newcommand \g {\gamma}

\newcommand \si {\sigma}

\newcommand \x {\cdot}

\newcommand \A {\alpha}
\newcommand \B {\beta}

\newcommand \lc {\langle}
\newcommand \rc {\rangle}
\newcommand \prt {\partial}

\newcommand \bvec{\left( \begin{array}{c} }
\newcommand \evec{\end{array} \right)}

\newcommand \eg {{\it e.g.}}  
\newcommand \bea{\begin{eqnarray} }
\newcommand \eea{\end{eqnarray} }
\newcommand \nn {\nonumber}
\newcommand {\be} {\begin{equation}}
\newcommand {\ee} {\end{equation}}
\newcommand {\epem} {$e^+ e^-$}

\begin{document}

\title{Evolution of the parton dihadron fragmentation functions}

\author{A. Majumder}
\affiliation{Nuclear Science Division, 
Lawrence Berkeley National Laboratory\\
1 Cyclotron road, Berkeley, CA 94720}

\author{Xin-Nian Wang}
\affiliation{Nuclear Science Division, 
Lawrence Berkeley National Laboratory\\
1 Cyclotron road, Berkeley, CA 94720}

\date{ \today}

\begin{abstract} 
Quark and gluon parton dihadron fragmentation functions and 
their evolution are studied in the process of $e^+e^-$ annihilation. 
We provide definitions of such dihadron fragmentation functions 
in terms of parton matrix elements and derive the momentum sum rules 
and their connection to single hadron fragmentation functions.
We parameterize results from the Lund Monte Carlo model JETSET as
the initial conditions for the parton dihadron fragmentation functions
at the scale $Q_0^2=2$ GeV$^2$. The evolution equations for the 
quark and gluon fragmentation functions are solved numerically and 
the results at different higher scales $Q^2$ agree well 
with JETSET results. The importance of the input from the single 
fragmentation functions is pointed out.  
\end{abstract}


\preprint{LBNL-52689}

\maketitle


\section{INTRODUCTION}


In the analysis of particle production from jets in \epem annihilation 
or in $p+p(\bar{p})$ collisions, a primary observable is the single 
particle inclusive cross section.
Within perturbative QCD (pQCD) and at leading-twist, 
the single particle inclusive cross 
section in \epem annihilation can be proved to factorize into a 
calculable perturbative hard partonic part and a nonperturbative
single inclusive fragmentation function \cite{col89}. In elementary 
particle collisions it is also possible to measure 
multiple particle production cross sections. This has led to the 
construction and study of  multi-particle observables such as event 
shapes \cite{Ellis:1980nc,Catani:1991kz}. These offer further insight 
into the substructure of jets produced in 
high energy particle collisions. In the analysis of jets produced in 
high energy heavy-ion collisions or in semi-inclusive 
deeply inelastic scattering (DIS) off large nuclei, however, the latter 
analysis is quite infeasible. Along with the single inclusive 
cross section another measurable quantity  which 
may offer insight into the modification of jet properties in a 
medium are two high momentum particle correlations. 

Dihadron correlations have indeed been measured recently 
in a variety of experiments. In the study of jet suppression 
in heavy-ion collisions, correlations between two high $p_T$ hadrons 
in azimuthal angle are used to study the medium modification of 
jet structure in heavy-ion collisions at the Relativistic Heavy-ion
Collider (RHIC) \cite{adl03,Adler:2004zd}. While the back-to-back correlations 
are suppressed in central $Au+Au$ collisions, indicating large
parton energy loss in the dense medium, the same-side 
correlations remain approximately the same as in $p+p$ and $d+Au$ collisions.
This is to be contrasted with the large suppression observed in 
single inclusive spectra \cite{highpt}. 
Two hadron correlations within the same jet are also measured in 
DIS off various nuclei at the HERMES experiment \cite{din04}. 
In such experiments the dihadron correlations surprisingly display 
minimal variation with the choice of nuclear target, even though the 
single inclusive production of leading hadrons are significantly 
suppressed with increasing atomic number of the nuclear 
target \cite{Airapetian:2000ks}.  Given the experimental kinematics of both 
experiments, this may be considered as an indication of parton 
hadronization outside the medium. However, since the
same-side correlation corresponds to the two-hadron distribution
within a single jet, the observed phenomenon is highly nontrivial.

To study systematically such phenomena, an extension of the single 
inclusive fragmentation formalism of QCD is required, to 
include correlations between pairs of particles produced in 
the same jet. Toward this end, we had proposed dihadron 
fragmentation functions in a recent paper \cite{maj04a}. 
In that effort, the double differential cross section for the 
same-side production of two hadrons in the \epem  annihilation 
was factorized into the well known hard cross section for
$e^+ + e^- \rightarrow \gamma^* \rightarrow q\bar{q}$ and
the quark dihadron fragmentation functions. The essential purpose 
of this fragmentation function is to measure the distribution of 
hadron pairs produced in the fragmentation of a hard quark. 

As shown in the case of single hadron fragmentation 
functions \cite{guowang}, the medium modification of the 
fragmentation function due to multiple
scattering and induced gluon radiation closely resembles 
that of radiative corrections due to gluon bremsstrahlung in 
vacuum. Therefore, it is important to understand first
the QCD evolution of the dihadron correlations 
in quark and gluon jets in vacuum.
In the last study \cite{maj04a}, we had concentrated on the
definition of the dihadron fragmentation functions and derived the
Dokshitzer-Gribov-Lipatov-Altarelli-Parisi (DGLAP) evolution 
equations \cite{gri72,dok77b,alt77} for the non-singlet quark 
dihadron fragmentation function. In this paper, we will consider
singlet quark and gluon dihadron fragmentation functions which
will be directly relevant to the study of medium modification of
jets in high-energy heavy-ion and DIS collisions. In addition,
we will explore momentum sum rules for the dihadron fragmentation 
functions and their relationship to single hadron fragmentation
functions.

In the factorized form for the inclusive hadron production cross
section, the short distance parton cross section can be computed 
order by order as a series in $\alpha_s(Q^2)$ for reactions with 
momentum transfer much higher above $\Lambda_{QCD}$.
The long distance objects or the inclusive $n$-hadron fragmentation 
functions contain the non-perturbative information of parton 
hadronization \cite{col89,maj04a}. These fragmentation functions can be  
defined in an operator formalism \cite{mue78,maj04b} and hence are valid 
beyond the perturbative theory. They however cannot 
be calculated perturbatively and have to be instead inferred from 
experiments. Since the definitions of these functions are universal 
or process independent, once measured in one 
process, {\it e.g.} $e^+e^-$ annihilation, they can be applied to 
another, {\it e.g.} deep inelastic scattering or $p+p$ collisions, 
and therein lies the predictive power of pQCD.
Yet another predictive power of pQCD rests in the fact that once these 
fragmentation functions are measured or given at one energy scale, 
they can be predicted for all other energy scales via the DGLAP
evolution equations. 

While hadronization of the out going quark and gluon jets is a 
non-perturbative phenomena, inclusive hadron production cross 
sections in $e^+e^-$ collisions have turned out to be one of 
the many successful predictions of perturbative 
QCD \cite{dok77b,ste77,fie78}. To the best of our knowledge, however,
measurements of dihadron fragmentation functions 
have not been performed in \epem experiments. In the
numerical study of the nonsinglet quark dihadron fragmentation 
functions in Ref.~\cite{maj04a}, we used a simple ansatz for the 
initial condition as $D_{NS}(z_1,z_2) = D(z_1)\times D(z_2)$, which
at best was just a guess and, as we will show later, differ 
significantly from the inherent hadron correlations in a single jet.
To facilitate a more accurate numerical
study of the singlet quark and gluon dihadron fragmentation
functions in this paper, we will parameterize the results of 
the Lund Monte Carlo model JETSET \cite{sjo95} which has successfully
described many aspects of jet fragmentation in both \epem and
DIS experiments.

The remaining sections of this paper are organized as follows. 
In Sec. II we review the definition of the double hadron
fragmentation function. We outline how 
such a function may be isolated in the expression for 
a double differential inclusive cross section.
 In Sec. III we derive and discuss various sum 
rules that are obeyed by the dihadron fragmentation functions and
their connection to single hadron fragmentation functions.
In Sec. IV we outline the  derivation of the 
DGLAP \cite{gri72,dok77b,alt77} evolution equations for the quark
and gluon dihadron fragmentation functions. 
The initial conditions for the dihadron fragmentation functions 
are extracted from JETSET at a scale $Q_0$ in Sec. V. They are
then evolved numerically via the DGLAP evolution equations to different
scales $Q$ and compared to the result form JETSET.
Finally in Sec. VI we discuss the results of our calculation and 
present our conclusions.


\section{Dihadron fragmentation functions}


In this section, we will review the definition and properties of 
dihadron fragmentation functions in the the semi-inclusive process
$e^+ + e^- \ra \gamma^* \ra h_1 + h_2 + X$
via single jet fragmentation where two hadrons are identified.
In most cases we will be concerned with back-to-back quark 
and antiquark jets. Study of the gluon fragmentation function, however, 
will involve three jet events. In both cases, our focus will always 
be on two hadrons produced in the same jet.

The cross section for the process of \epem annihilation into hadrons may 
be expressed as 
\bea 
\si &=& \frac{1}{2s} \frac{e^4}{4 (q^2)^2} \sum_{S_{had}}
%
%
\kd^4(k_1 + k_2 - P_S) \nn \\
&\times&\mal_{\mu \nu } 
\lc 0 | J^\mu(0) | S_{had} \rc \lc S_{had} | J^{\nu} (0) | 0\rc \nn \\
&\equiv & \frac{e^4}{2s q^4} 
\frac{\mal_{\mu \nu} W^{\mu\nu}}{4}, \label{s_w}
\eea 
where $J^\mu=\sum_q \bar{\psi}_q\gamma^\mu \psi_q$ is the hadronic 
electromagnetic current, $\mal_{\mu \nu}$ the leptonic tensor 
and $W^{\mu \nu}$ the hadronic tensor. The four momentum of the 
virtual photon is $q = k_1 + k_2 \equiv (Q,0,0,0)$ and the Mandelstam
variable $s = q^2 = Q^2$. In the remaining, the sum over $S_{had}$
will include both the sum over the complete set of states and the 
phase space integration $\Pi_{f\in S_{had}} d^3 p_f / 2 E_f (2 \pi )^3$
and $P_S=\sum_f p_f$.

\begin{figure}[htb!]
  \resizebox{3.0in}{3.0in}{\includegraphics[1in,1in][9in,9in]{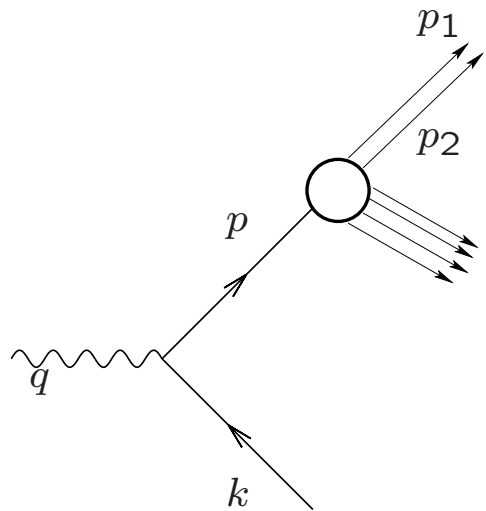}} 
    \caption{The leading order Feynman diagram contributing to the 
    double inclusive fragmentation function.}
    \label{fig1}
\end{figure}

The definition and factorization of dihadron fragmentation functions 
involve identifying two hadrons with nearly parallel momenta $p_1$ 
and $p_2$ among hadronic states along the direction of one of the partons 
and replacing the remaining sum over hadronic states with a sum over 
the rest of all partonic states (see Fig.~\ref{fig1}). This is followed by 
an extraction of the leading twist component (see Ref.~\cite{maj04a} for 
details). Differentiating the total inclusive cross section with respect to 
the forward momentum fractions of the  two hadrons \tie, $z_1,z_2$, the 
leading order (LO) factorized form of the double differential cross 
section may be expressed as
\bea
\frac{d^2 \si}{dz_1 dz_2} = \sum_q \si_0^{q\bar{q}} 
\left[ D_q^{h_1 h_2} (z_1,z_2) + 
D_{\bar{q}}^{h_1 h_2} (z_1,z_2) 
\right],
\label{LO_Dz1z2}
\eea
where $\si_0^{q\bar{q}}$ is the total LO hard cross section for an 
\epem pair to annihilate into hadrons. The two functions 
$D_{q}^{h_1 h_2} (z_1,z_2) $  and $D_{\bar{q}}^{h_1 h_2} (z_1,z_2)$  
represent the dihadron fragmentation functions for a quark and an antiquark.
The quark dihadron fragmentation function in light-cone gauge ($A^+ = 0$ gauge)
is obtained as,
\bea
D_q^{h_1,h_2}(z_1,z_2) &=& \frac{z^4_h}{4z_1z_2} 
\int \frac{d^2q_\perp}{4(2\pi)^3} 
\int \frac{d^4 p}{(2\pi)^4} \nn \\ 
&\times& T_q(p;p_1,p_2) 
\kd \left( z_h - \frac{p_h^+}{p^+}  \right). \label{dihad_qrk}
\eea
The forward momentum fractions of the identified hadrons $z_1,z_2$ are
$z_1 = p_1^+/p^+$ and $z_2 = p_2^+/p^+$. 
The momentum $p_h$ represents the sum of the hadronic momenta 
\tie, $p_h = p_1 + p_2$. Henceforth, the direction identified by 
$\vec{p}_h$ will be considered the same as the direction of the jet.
The transverse spread of the two hadrons around the jet direction is 
indicated by the relative transverse 
momentum  $\vec{q}_\perp = \vec{p}_{1 \perp} - \vec{p}_{2 \perp}$. 
As is indicated by the final $\delta$-function in the above equation, 
the momentum fraction $z_h = z_1 + z_2$. 
The matrix element of the operator product $T_{q}(p;z_1,z_2)$ is 
given as (in the light-cone gauge)
\bea
T_q(p;p_1,p_2)
&=&  {\rm Tr} \Bigg[ \frac{\g^+}{2p_h^+} 
\int d^4 x e^{i p \x fx} \sum_{S - 2} \nn \\
& & \hspace{-1.0in} \times 
\lc 0 | \psi_q (x) | p_1, p_2, S-2 \rc  
\lc p_1, p_2, S-2 | \bar{\psi}_q (0) | 0 \rc \Bigg].
\label{t_q}
\eea
The above expression for the dihadron fragmentation functions 
includes the integration over the transverse momenta $q_\perp$. 
As a result the angular correlation between the 
detected hadrons is also integrated over in such a definition.
In order to observe such correlation a differentiation of the 
dihadron fragmentation function with respect to the transverse angle 
must be carried out. In this paper, we will continue to focus on the 
integrated fragmentation function.

\begin{figure}[htb!]
  \resizebox{3.in}{3.0in}{\includegraphics[0in,0in][6in,6in]{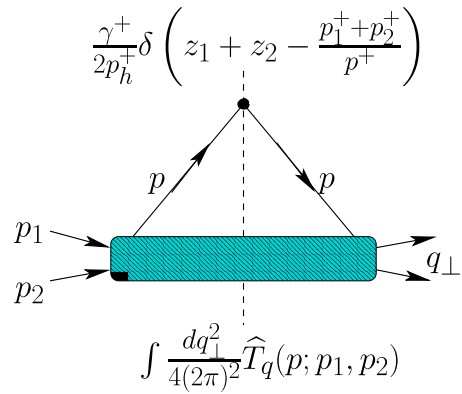}} 
    \caption{The cut-vertex representation of the quark dihadron fragmentation
    function. }
    \label{fig2}
\end{figure}

The above definition for the dihadron fragmentation function also 
admits a simple diagrammatic interpretation in terms of the 
cut-vertex method of Mueller \cite{mue78}. The diagrammatic 
rule in terms 
of the quark cut-vertex is outlined in Fig.~\ref{fig2}. 
The product of these rules along with the factor $z^4_h/(4z_1z_2)$ is 
the integrated dihadron fragmentation function.

The gluon dihadron fragmentation function can be similarly constructed 
by identifying two hadrons moving with almost parallel momenta 
in the direction of the outgoing gluon in a three jet event
in \epem annihilation processes.
Factorizing the hard cross section from the soft 
matrix element one obtains the 
gluon dihadron fragmentation function at leading twist as 
(see Ref.~\cite{maj04a} for details),
\bea 
D_g(z_1,z_2) &=& \frac{z^3_h}{2 z_1 z_2 }  \int \frac{dq_\perp^2}{8(2\pi)^2}
\int \frac{d^4l}{(2\pi)^4} \nn \\
&\times&
\kd \left( z_h - \frac{p_h^+}{l^+} \right) T_g (l;p_1,p_2).
\label{dihad_glu}
\eea
In the above equation, the meanings of various momenta and 
forward momentum fractions are the same as for the quark dihadron
fragmentation function. The gluon overlap matrix 
element $T_g (l;p_1,p_2)$  is defined as
\bea 
T_g (l;p_1,p_2) &=& \int d^4 x e^{i l \x x} \sum_{S-2} 
\lc 0 | A^a_\mu (x) | p_1,p_2,S-2 \rc \nn \\
&\times &\lc p_1,p_2,S-2 | A^b_\nu(0) | 0 \rc 
\frac{\kd^{ab} d^{\mu \nu}(l)}{16}, \label{glue_matrix}
\eea
where $d^{\mu \nu}(l)$ is the gluon's polarization tensor in the 
light-cone gauge and sum over the color indices of the gluon field 
is implied.

\begin{figure}[htb!]
  \resizebox{3.0in}{3.0in}{\includegraphics[0in,0in][6in,6in]{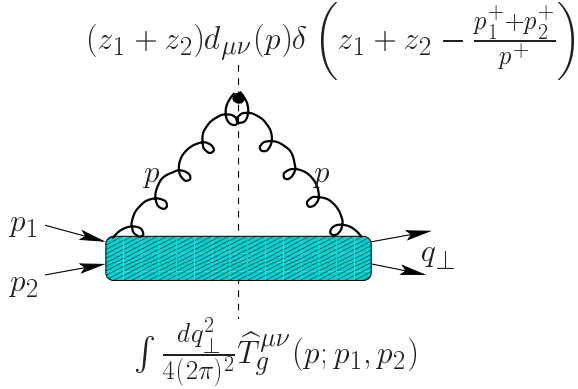}} 
    \caption{The cut-vertex representation of the gluon dihadron fragmentation
    function. }
    \label{fig3}
\end{figure}

The gluon dihadron fragmentation function also 
admits a simple diagrammatic interpretation in terms of the 
cut-vertex method. The diagrammatic 
rule in terms 
of the gluon cut-vertex is outlined in Fig.~\ref{fig3}. 
The product of these rules along with the factor $z^2_h/(2z_1z_2)$ is 
the integrated dihadron fragmentation function. 


\section{Sum rules}


One of the many interesting properties 
obeyed by the single inclusive fragmentation 
functions are the various sum rules. 
Primary among them are the momentum sum rules
\bea
\sum_{h} \int dz  z D_q^h (z) = \sum_{h} \int dz  z D_g^h (z) = 1. 
\eea
Here the single inclusive fragmentation functions are 
defined as \cite{mue78,mue81,osb03},
\bea
D_{q}^h (z) 
&=& \frac{z^3}{2} \int \frac{d^4 p}{(2\pi)^4}
\kd \bigg( z - \frac{p_h^+}{p^+} \bigg) T_{q} (p,p_h);
\label{dz1} \\
D_{g}^h (z) 
&=& z^2 \int \frac{d^4 l}{(2\pi)^4}
\kd \bigg( z - \frac{p_h^+}{l^+} \bigg) T_{g} (p,p_h),
\label{dz2}
\eea
in terms of parton matrix elements
\bea 
T_q (p,p_h) &=& \mbox{Tr} \bigg[ \frac{\gamma^+}{2p_h^+} 
\int d^4 x \sum_{S-1} 
\lc 0 | \psi_q (0) |p_h, S-1 \rc \nn \\
&\times& \lc \ p_h, S-1 | \bar{\psi}_q (x) | 0 \rc  e^{i p\x x} \bigg];
\label{dz1_op} \\
T_g(l,p_h) &=& 
\int d^4 x \sum_{S-1} 
\lc 0 | A^a_\mu (x) |p_h, S-1 \rc \nn \\
&\times& \lc \ p_h, S-1 | A^b_\nu(0) | 0 \rc  
\frac{\kd^{ab} d^{\mu \nu}(l)}{16} e^{i l\x x}.
\label{dz2_op}
\eea
They can be interpreted as single inclusive hadron multiplicity 
distributions within fractional momentum $z_h$ and $z_h+dz_h$ 
from parton fragmentation.
Similarly, one can derive the momentum sum rules for dihadron
fragmentation functions. Furthermore, one can also derive
sum rules that relate dihadron fragmentation functions to
single fragmentation functions.

Given $\vec{p}_{\perp h}=\vec{p}_{\perp 1}+\vec{p}_{\perp 2}$ 
and $\vec{q}_{\perp}=\vec{p}_{\perp 1}-\vec{p}_{\perp 2}$, we have 
assumed that the total transverse momentum of the two 
hadrons $\vec{p}_{\perp h}$ is parallel to the parton's transverse 
momentum in the definition of parton dihadron fragmentation functions 
in Eqs.~(\ref{dihad_qrk}) and (\ref{dihad_glu}). Similarly, if we
consider single hadron fragmentation, the hadron transverse momentum
$\vec{p}_{\perp,1}$ is also parallel to the partons transverse momentum.
We should then have the following identity,
\begin{equation}
\int \frac{d^2p_{\perp h}}{2\pi z_h^2 p_q^{+2}}
=\int \frac{d^2p_{\perp 1}}{2\pi z_1^2 p^{+2}_q}=1, \label{iden1}
\end{equation}
which is just integration over the angle of the initial parton.
Using the above identity, one can recast the momentum integration as,
\begin{equation}
\frac{dz_2}{z_2}d^2p_{\perp h}\frac{d^2q_\perp}{4(2\pi)^3}
=2 d^2p_{\perp 1}\frac{d^3p_2}{2E_2(2\pi)^3}. \label{iden2}
\end{equation}
Here, the external quark momentum is $p_q^+$ and hadrons momenta are
$p_1^+=z_1p_q^+$ and $p_2^+=z_2p_q^+$. One can also 
rewrite the $\delta$-function 
$\delta(z_h-p_h^+/p^+)=(z_1/z_h)\delta(z_1-p_1^+/p^+)$
in the definition of the dihadron fragmentation functions.

Since a dihadron fragmentation function is essentially a two-hadron
multiplicity distribution, its sum rules must involve the information
of the number of hadron pairs. An example of this is
\begin{equation}
\sum_{h_1,h_2} \int dz_1 dz_2 D_{q,g}^{h_1 h_2}(z_1,z_2)
=\langle N(N-1)\rangle, \label{2moment}
\end{equation}
where $\langle N(N-1)\rangle$ is the second cumulant moment of 
the multiplicity distribution and
\begin{equation}
\langle N\rangle =\sum_h \int dz D_{q,g}^h(z) \label{1moment}
\end{equation}
is the mean multiplicity from parton fragmentation. Inside the
parton matrix elements of the fragmentation function, one can
rewrite
\begin{eqnarray}
& &\sum_{h_2,S-2} \int \frac{d^3p_2}{2 E_2 (2\pi)^3} 
|p_1, p_2, S-2 \rc \lc \ p_1, p_2, S-2 | \nn \\
&=&\sum_{h_2} \int \frac{d^3p_2}{2 E_2 (2\pi)^3}
\hat{a}^\dagger_{h_1}(p_1)\hat{a}^\dagger_{h_2}(p_2)
\hat{a}_{h_2}(p_2)\hat{a}_{h_1}(p_1), \nn \\
&=&\hat{a}^\dagger_{h_1}(p_1) \hat{a}_{h_1}(p_1) (\hat{N}-1) \\
\end{eqnarray}
where $\hat{a}^\dagger_h$ and $\hat{a}_h$ are creation and annihilation
operators for hadron $h$, and
\begin{equation}
\hat{N}=\sum_{h} \int \frac{d^3p}{2 E (2\pi)^3}
\hat{a}^\dagger_{h}(p) \hat{a}_{h}(p),
\end{equation}
is the total multiplicity operator. Here we have assumed hadrons
as bosons. Using similar expression,
\begin{equation}
\sum_{S-1} |p_1, S-1 \rc \lc \ p_1, S-1 | 
=\hat{a}^\dagger_{h_1}(p_1) \hat{a}_{h_1}(p_1),
\end{equation}
for single inclusive states, one can effectively have the identity,
\begin{eqnarray}
& & \sum_{h_2,S-2} \int \frac{d^3p_2}{2 E_2 (2\pi)^3} 
|p_1, p_2, S-2 \rc \lc \ p_1, p_2, S-2 | \nn \\
&=&\frac{\langle N(N-1)\rangle}{\langle N\rangle}
\sum_{S-1} |p_1, S-1 \rc \lc \ p_1, S-1 |. \label{iden3}
\end{eqnarray}
Using Eqs.~(\ref{iden1}),(\ref{iden2}) and (\ref{iden3}),
one can obtain the following relationship,
\begin{equation}
\sum_{h_2}\int dz_2 D_{q,g}^{h_1 h_2}(z_1,z_2)
=\frac{\langle N(N-1)\rangle}{\langle N\rangle} D_{q,g}^{h_1}(z_1),
\end{equation}
between dihadron and single hadron fragmentation functions. Using
the momentum sum rules for the single hadron fragmentation functions, one
has the following momentum sum rule for dihadron fragmentation
functions,
\begin{equation}
\sum_{h_1,h_2}\int dz_1 dz_2 \frac{z_1+z_2}{2} 
D_{q,g}^{h_1 h_2}(z_1,z_2)=\frac{\langle N(N-1)\rangle}{\langle N\rangle}.
\end{equation}
Note that both $\langle N\rangle$ and $\langle N(N-1)\rangle$
in Eqs.~(\ref{2moment}) and (\ref{1moment}) are not infrared safe
and therefore not well defined
in the collinear approximation for the fragmentation functions.
Consequently, these sum rules for dihadron fragmentation functions
are also not well defined in collinear approximation. They, however,
may provide useful phenomenological constraints for practical
modeling of these fragmentation functions where collinear
approximation might not be required.


\section{Dglap evolution}


One of the many successes of the factorized pQCD 
is the prediction of the scaling violation
of the single inclusive fragmentation functions via the DGLAP 
equations. If the fragmentation functions are measured at 
an initial large scale $Q_0 >> \Lambda_{QCD}$, 
they can be predicted at any higher scale $Q$. In terms of 
the physical pictures afforded by the parton model, one 
imagines that the initiating quark loses its large virtuality 
through the radiation of multiple soft gluons and finally 
fragment into hadrons. Such a picture is validated 
within field theory by considering the leading twist contributions
of higher order diagrams in \epem annihilation. The leading 
log (LL) contributions from all such diagrams are resummed and 
a set of coupled differential equations for the rate of change 
of the fragmentation functions with the momentum scale is derived.

\begin{figure}[htb!]
  \resizebox{3in}{3in}{\includegraphics[-0.5in,-0.5in][5in,5in]{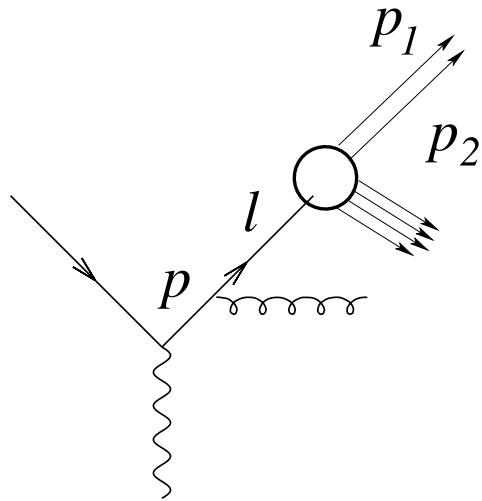}} 
    \caption{The contribution from the quark fragmentation
      to the NLO correction of the 
    quark dihadron fragmentation function. }
    \label{fig4}
\end{figure}

\begin{figure}[htb!]
  \resizebox{3in}{3in}{\includegraphics[0in,0in][5in,4in]{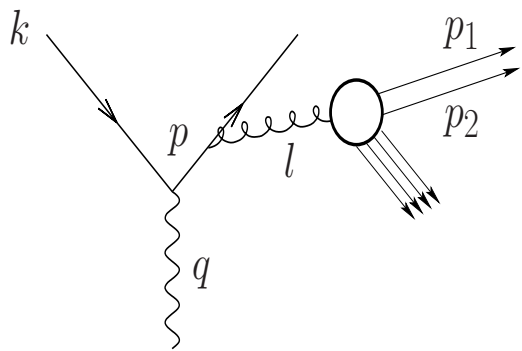}} 
    \caption{The contribution of gluon fragmentation to the NLO 
    correction of the quark dihadron fragmentation function. }
    \label{fig5}
\end{figure}

\begin{figure}[htb!]
  \resizebox{4in}{3.5in}{\includegraphics[0in,0in][6in,5in]{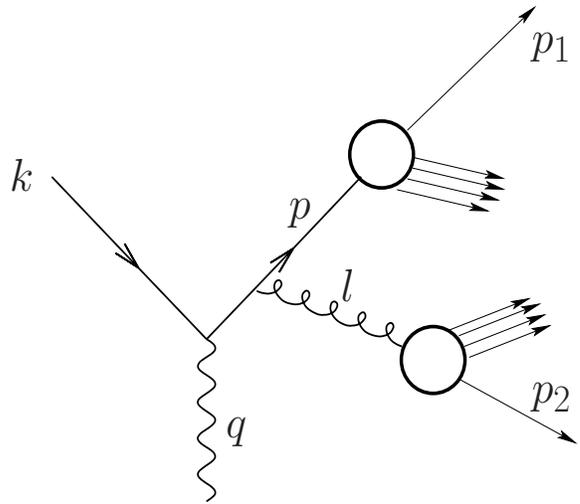}} 
    \caption{The mixed contribution to the 
NLO correction of the quark dihadron 
fragmentation function.}
    \label{fig6}
\end{figure}

In Ref.~\cite{maj04a}, a QCD evolution equation for the
non-singlet quark dihadron fragmentation function in an operator 
formalism was derived. The resulting evolution equation also admits a 
simple physical interpretation in terms of the diagrams shown in 
Figs. \ref{fig4}, \ref{fig5}, \ref{fig6}. The reader will note the 
new contribution from the diagram indicated in Fig. \ref{fig6}: 
the two detected hadrons emerging from the fragmentation of 
two separate partons. This contribution will turn out to be
essential in determining the QCD evolution of dihadron fragmentation
functions. 

The resulting DGLAP evolution equation for the quark 
dihadron fragmentation function is given as 

\begin{widetext}
\bea 
\frac{\prt D_q^{h_1 h_2} (z_1,z_2,Q^2)}{\prt \log{Q^2}} &=& 
\frac{\A_s}{2\pi} \Bigg[ \int_{z_1 + z_2}^1 \frac{dy}{y^2} P_{q\ra q g} (y)  
D_q^{h_1 h_2} \left( \frac{z_1}{y}, \frac{z_2}{y}, Q^2 \right) \nn \\
&+&  \int_{z_1}^{1-z_2} \frac{dy}{y(1-y)} \hat{P}_{q \ra q g}(y) 
D_{q}^{h_1} \left(\frac{z_1}{y}, Q^2 \right) 
D_g^{h_2} \left( \frac{z_2}{1-y},Q^2 \right) \nn \\
&+& \int_{z_2}^{1-z_1} \frac{dy}{y(1-y)} \hat{P}_{q \ra q g}(y) 
D_{q}^{h_2} \left(\frac{z_2}{y}, Q^2 \right) 
D_g^{h_1} \left( \frac{z_1}{1-y},Q^2 \right)\nn \\
&+& \int_{z_1 + z_2}^1 \frac{dy}{y^2} P_{q\ra g q} (y)  
D_g^{h_1 h_2} \left( \frac{z_1}{y}, \frac{z_2}{y}, Q^2 \right) \Bigg].
\label{q_dglap}
\eea
\end{widetext}
Where $D_q^{h} (z, Q^2)$ and $D_g^h (z,Q^2)$ are the single inclusive 
quark and gluon fragmentation functions, and $D_g^{h_1 h_2} (z_1,z_2,Q^2)$ 
is the 
gluon dihadron fragmentation function [see Eq. (\ref{dihad_glu})].  
In the above equation $P_{q \ra qg}(y)$ is the splitting function for a quark 
to radiate off a gluon and keep a fraction $y$ of its initial forward 
light-cone momentum. It is identical to the splitting function kernel 
of the DGLAP equation for the single fragmentation function, \tie,
\bea
P_{q \ra qg} (y) = C_F \left( \frac{1+y^2}{1-y} \right)_+.
\eea
The subscript `+' indicates that the negative virtual correction has been 
added within the splitting function. 
The splitting function in the second line of Eq. (\ref{q_dglap}),
$\hat{P}_{q\ra qg}$, is identical to the above equation except 
that it lacks the negative virtual correction. The last splitting 
function is the probability for a quark to radiate off a gluon 
with a fraction $y$ of its forward momentum and has an expression 
identical to the case for the single fragmentation case. It should 
be pointed out in passing that $P_{q\ra g q} (y) $ also has no virtual 
correction and thus no negative contribution. In the evolution of the dihadron 
fragmentation function the sole negative contribution arises from the virtual 
piece in $P_{q \ra qg}(y)$; all other contributions are positive. Another 
difference from the case of the single inclusive fragmentation function are 
the measures $1/y^2$ and $1/y(1-y)$. These can be understood on the basis that 
the origin of the dihadron fragmentation function lies in the evaluation 
of  a double differential inclusive cross section (see Sec. II of 
Ref. \cite{maj04a}). We also point out that Eq. (\ref{q_dglap})  
may also be derived in the cut-vertex formalism via the 
renormalization of the bare cut-vertex presented in Fig. \ref{fig2}.

The evolution of the gluon dihadron fragmentation function may be derived 
in the cut-vertex formalism via the renormalization of the bare cut-vertex 
shown in Fig. \ref{fig3}. 
It should be pointed out that the 
evolution equations for the quark and 
gluon dihadron fragmentation functions may be motivated using parton 
model arguments such as those of Refs. \cite{fie78,fie95}. 
The evolution equation for the gluon dihadron fragmentation functions
now includes four pieces: the gluon may split into a 
quark-antiquark pair or into two gluons. The detected hadrons 
may both emanate from the same parton (quark, antiquark or gluon) 
or from the two different partons resulting from the split, \tie,  
\begin{widetext}
\bea
\frac{\prt D_g^{h_1 h_2} (z_1,z_2,Q^2)} {\prt \log{Q^2}} &=& 
\frac{\A_s}{2\pi} \Bigg[ \int_{z_1 + z_2}^1 
\frac{dy}{y^2} 2 n_f P_{g \ra q \bar{q}} (y)  
D_q^{h_1 h_2} \left( \frac{z_1}{y}, \frac{z_2}{y}, Q^2 \right)  \nn \\
&+&  \int_{z_1}^{1-z_2} \frac{dy} {y(1-y)} n_f  P_{g \ra q \bar{q}}(y) 
D_{q}^{h_1} \left(\frac{z_1} {y}, Q^2 \right) 
D_{\bar{q}}^{h_2} \left( \frac{z_2} {1-y},Q^2 \right)  \nn \\
&+&  \int_{z_2}^{1-z_1} \frac{dy}{y(1-y)} n_f  P_{g \ra q \bar{q}}(y) 
D_{q}^{h_2} \left(\frac{z_2}{y}, Q^2 \right) 
D_{\bar{q}}^{h_1} \left( \frac{z_1}{1-y},Q^2 \right)  \nn \\
&+& \int_{z_1 + z_2}^1 \frac{dy}{y^2} P_{g\ra g g} (y)  
D_g^{h_1 h_2} \left( \frac{z_1}{y}, \frac{z_2}{y}, Q^2 \right) \nn \\
&+&  \int_{z_1}^{1-z_2} \frac{dy}{y(1-y)} \hat{P}_{g \ra g g }(y) 
D_{g}^{h_1} \left(\frac{z_1}{y}, Q^2 \right) 
D_{g}^{h_2} \left( \frac{z_2}{1-y},Q^2 \right)  \Bigg]. 
\label{g_dglap}
\eea
\end{widetext}
In the above equation, $n_f$ is the number of flavors of quarks assumed.
In the remainder of this paper we will assume $n_f$ to be 3. Henceforth, 
the quark and antiquark fragmentation functions will not be distinguished; 
in both cases the singlet fragmentation function will be assumed \tie,
$D_q = D_{\bar{q}} = ( D_q + D_{\bar{q}})/2 $. 
The splitting 
function $P_{g \ra q \bar{q}}$ is the probability for the initial gluon to 
decay into a quark-antiquark pair with the quark carrying 
a fraction $y$ of the 
forward momentum of the gluon. It is given as, 
\bea
P_{g \ra q \bar{q}} &=& T_F (y^2 + (1-y)^2),
\eea
where $T_F$ is the Casimir. It may be noted that interchanging 
quark and antiquark may also be achieved by the switch $y \ra 1-y$.
In Eq.~(\ref{g_dglap}), $P_{g\ra g g} (y)$ is the probability for the 
gluon to split into two gluons, and has the expression
\bea
P_{g\ra g g} (y) &=& 2 C_A \Bigg[ \frac{y}{(1-y)_+} 
+ \frac{1-y}{y} + y(1-y) \Bigg] \nn \\
&+& \kd(1-y) \Bigg[ \frac{11}{6} C_A - \frac{2}{3} n_f T_F \Bigg] .
\eea 
The presence of the `+'-function as well as the $\delta$-function 
is the result of virtual contributions from gluon and quark-antiquark 
loops. The final splitting function 
in Eq. (\ref{g_dglap}) is essentially the above splitting function 
without the virtual corrections, \tie,
\bea
\hat{P}_{g\ra g g} (y) &=& 2 C_A \Bigg[ \frac{y}{(1-y)} 
+ \frac{1-y}{y} + y(1-y) \Bigg],
\eea 
since there are no virtual corrections to the independent 
fragmentation contributions at the same order. 
As hadrons with finite momentum fractions originating from both gluons are 
detected, the range of values of the intermediate momentum fraction $y$ may 
approach neither 0 nor 1. Therefore, there is no infrared divergence
even without cancellation by virtual corrections. 
As one can see now there exist two new contributions to 
the evolution of the gluon dihadron fragmentation function: from 
independent single fragmentation functions 
after a split to a quark-antiquark and to two gluons. Both these 
contributions are positive. This again will turn out to be an essential 
part in the equation that influences the QCD evolution of gluon
dihadron fragmentation functions.


\section{Results of evolution: comparison with event generators}


To date, measurements of single inclusive cross sections in 
\epem  collisions remain the primary set of data used in the 
parameterization of the fragmentation functions and testing their 
scaling violations. In many ways this provides an independent 
justification of the factorized pQCD approach to high energy collisions.
The absence of any initial state interactions makes \epem 
experiments ideal for baseline measurements of parton fragmentation
in vacuum,
while allowing for tractable calculations of their evolution. 
However, no such measurements have been performed for double 
inclusive cross sections. In the absence of such measurements and 
in the interest of simplicity we turn to Monte Carlo event generators 
for both the extraction of the initial conditions and for comparison
to the numerical results of the evolution equation for dihadron 
fragmentation functions.

Monte Carlo event generators such as JETSET \cite{sjo95} have 
enjoyed great success as simulators of \epem 
collision events. They provide reliable predictions 
not only for single inclusive measurements but also 
for many-particle observables such as event shape 
and inter-jet particle flows. Hence, it is reasonable to 
assume that two particle correlations extracted from 
a ``tuned'' event generator will closely mimic 
such correlations measured in real experiments. 
The events generated will be restricted to two-jet 
events for a measurement of the quark dihadron 
fragmentation function and to three-jet events for the 
gluon dihadron fragmentation function. Single inclusive
hadron fragmentation functions for both quarks and gluons 
will also be measured in the same set of events. 

In the extraction of the quark fragmentation functions three 
million dijet events distributed equally over three flavors 
were simulated using JETSET. 
The $\sqrt{s}$ of the reaction was set to 20 GeV.
In all such events the direction of the 
initiating quark and its initial virtuality were controlled. 
The virtuality of the initiating quark $Q$ sets the scale of the 
fragmentation function.
The forward light-cone momentum of the initiating quark was 
read off from the event list. In such experiments the 
single fragmentation function is defined as 
\bea
D_q(z) = \frac{1}{N_{evt}} \frac{dN(z)}{dz}.
\eea
In this paper, we restrict the flavor of the 
detected particles to $\pi^+$ and $\pi^-$. The 
variable $z=p_\pi^+/p_q^+$, where $p_q$ is the 
momentum of the fragmenting quark. The denominator 
$N_{evt}$ is the number of events, whereas 
$dN(z)$ represents the number of particles, 
over all events, that fall between $z$ and $z+dz$
of the momentum fraction. The dihadron 
fragmentation function is measured as
\bea
D_q(z_1,z_2) =\frac{1}{N_{evt}} \frac{d^2 P(z_1,z_2)}{dz_1 dz_2},
\eea
where, $d^2 P(z_1,z_2)$ represents the number of pairs of pions 
with momentum fractions between $(z_1,z_1+dz_1)$ and $(z_2,z_2+dz_2)$.
In constructing these functions, $z_1$ was always restricted 
to be larger than $z_2$.

\begin{figure}[htb!]
\resizebox{2.5in}{3.25in}{\includegraphics[1in,2in][7in,10in]{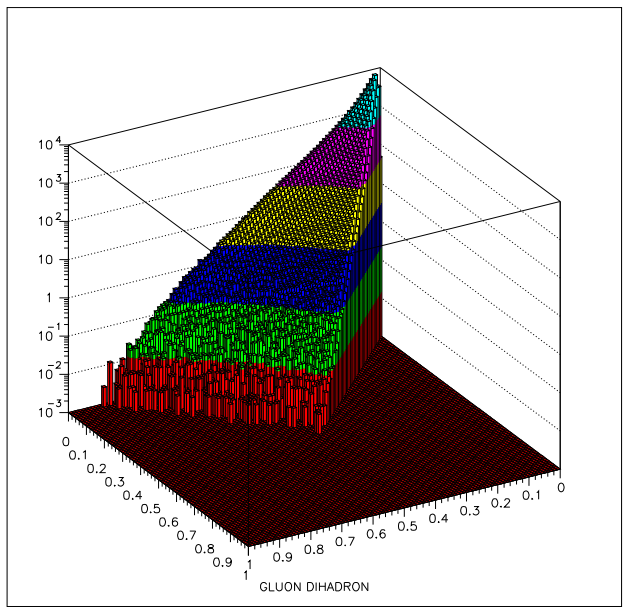}} 
    \caption{The gluon dihadron fragmentation function.}
    \label{fig7}
\end{figure}

For the extraction of the gluon fragmentation function, an identical 
procedure as above was carried out with the sole restriction, that all
events be three-jet events  and a large fraction of the energy 
be concentrated in the gluon. The latter requirement guarantees that 
events are predominantly composed of cases where the gluon and the 
quark antiquark pair are always contained in opposite hemispheres.
Tracking the final hadrons in the gluon's hemisphere lead to the 
construction of the gluon fragmentation function. It should be pointed
out that in the Lund model of fragmentation, which is the underlying 
fragmentation model in JETSET, an out going 
gluon is represented by a kink in
the fragmenting string which begins at the quark and terminates at the
antiquark. Perturbative showers within JETSET modify this picture and 
allow for multiple strings to form and fragment. A plot of the
gluon dihadron fragmentation function thus extracted is 
shown in Fig. \ref{fig7}.

\begin{figure}[htb!]
\resizebox{2.5in}{3.25in}{\includegraphics[1in,2in][7in,10in]{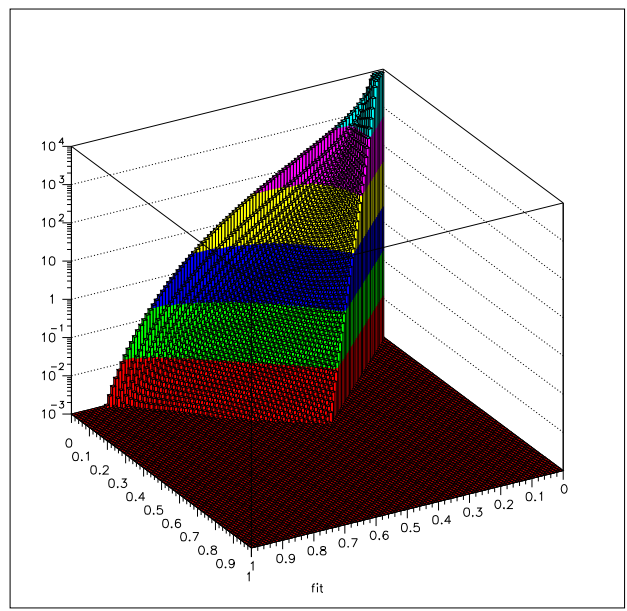}} 
    \caption{The result of $\chi^2$ fit of
    the function of Eq.~(\ref{fit}) to Fig.~\ref{fig7}.}
    \label{fig8}
\end{figure}

For the convenience of providing initial conditions to the evolution 
equations, the fragmentation function of Fig. \ref{fig7} is 
parametrized by fitting to a function of the type:
\bea
D(z_1,z_2) &=& N z_1^{\A_1} z_2^{\A_2} (z_1+z_2)^{\A_3} (1-z_1)^{\B_1}
(1-z_2)^{\B_2} \nn \\
&\times& (1-z_1-z_2)^{\B_3}. \label{fit}
\eea   
There are seven parameters in the above fit:
$N,\A_1,\A_2,\A_3,\B_1,\B_1,\B_3$.
The reader will note that the structure of the function 
is a simple generalization of the parameterization of the 
single fragmentation functions. The fit is carried out by the 
method of minimum logarithm of  $\chi^2$. The logarithm 
ensures that the fit is better at larger values of $z_1$ and $z_2$.
A plot of the final result of the fit for the gluon dihadron 
fragmentation function is shown in Fig.~\ref{fig8}.
One notes that the fit function of Eq.~(\ref{fit}) mimics the 
function closely except at very small $z_1$ and $z_2$. This is 
unimportant as in the evolution we will always restrict our 
attention away from very small $z_1$ and $z_2$. We will not explicitly
show the quark dihadron fragmentation function and its fit here. 
Suffice to say that the fit is even better for the 
quark, which, as is the case for the single fragmentation function,
generically has a harder spectrum in momentum fraction. Values for 
the various parameters of the fits to the quark and gluon fragmentation 
functions at the two different values of the $Q^2$ used in this paper 
are presented in table \ref{tab1}.

One unique feature of the DGLAP evolution equations for dihadron
fragmentation functions is that the equations couple dihadron to
single hadron fragmentation functions. The single hadron fragmentation
functions themselves evolve independently according to their own DGLAP
evolution equations. We find that the single hadron fragmentation
functions obtained from JETSET simulations can be described very well
by Binneswies-Kniehl-Kramer(BKK) parameterization \cite{bin95} 
of the actual experimental data. 
Therefore, in our numerical study of the evolution equations for
dihadron fragmentation functions, we will simply use the BKK 
parameterization for the single hadron fragmentation functions
and their evolution with the momentum scale $Q$.

\begin{figure}[htb!]
  \resizebox{3.2in}{4.1in}{\includegraphics[0.5in,1in][7.5in,10in]{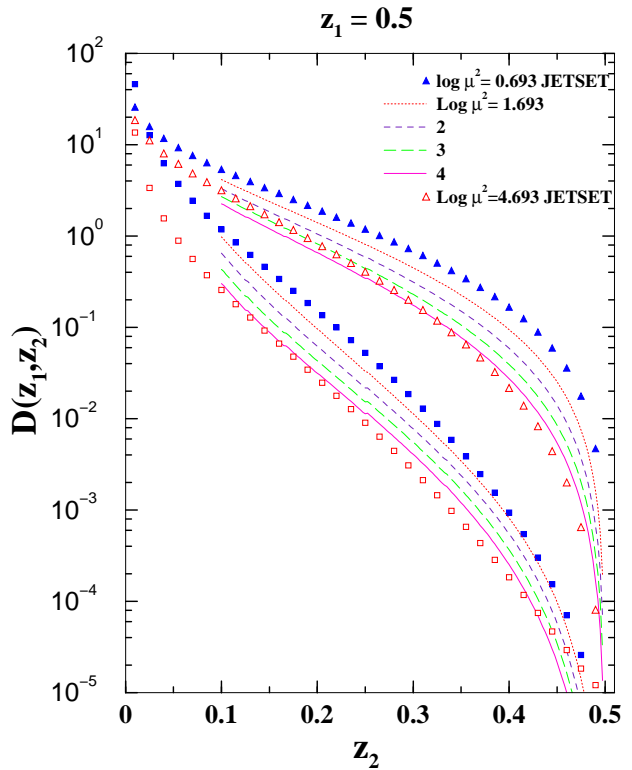}} 
    \caption{ Results of the evolution of the quark and gluon fragmentation 
    function ($D_q(z_1,z_2),D_g(z_1,z_2)$). In all cases $z_1$ is held fixed 
    at 0.5. }
    \label{fig9}
\end{figure}

The fit function shown in Fig.~\ref{fig8} as 
well as its analogue for the quark will provide 
the initial conditions to the differential equations outlined 
in Eqs.~(\ref{g_dglap}) and (\ref{q_dglap}) respectively. In both cases, 
the initial scale of the fragmentation functions is set to $Q_0^2 =
2$ GeV$^2$. This corresponds to $\log(Q_0^2) = 0.693$. 
These are shown as the filled triangles for the quarks
and the filled squares for the gluons in Fig~\ref{fig9}. 
We have chosen a fixed $z_1 = 0.5$ and let $z_2$ vary from 
$0$ to $1-z_1$, since several experiments that measure medium 
modification of the dihadron fragmentation functions, which 
we will discuss in separate studies, have similar kinematic range.
Note the orders of magnitude difference between the two distributions. 
This feature holds for most of the range of $z_1$ and $z_2$ except 
at very small momentum fractions where the gluon fragmentation 
function overtakes that of the quark.

Similar to the procedure carried out in Ref. \cite{maj04a} for
non-singlet quark dihadron fragmentation functions, results of the 
evolution will be presented in increments of $\log(Q^2) = 1$. As expected,
we note a softening of the spectrum with rising scale. We
terminate the evolution at $\log(Q^2) = 4.693$, corresponding 
to scale $Q^2 = 109$ GeV$^2$.
To compare the results of the evolution equations in 
Eqs.(\ref{q_dglap}) and (\ref{g_dglap}) to Monte Carlo event simulations, 
dihadron fragmentation functions at the highest scale are once 
again extracted from JETSET.
The number of events used to sample the fragmentation functions and the 
method of the fit remain identical to the case at lower $Q^2$ 
described above. The results are presented as open triangles for 
the quarks and open squares for the gluons. As shown they are in 
excellent agreement with the results of the evolution equations.   
This provides the most crucial test of the evolution equations
we have presented in this paper. 

\begin{table}[!htb] 
\begin{tabular}{| l | c | c | c | c | c | c | c |}
\hline
Parton & $N$ & $\A_1$ & $\A_2$ & $\A_3$ & $\B_1$ & $\B_2$ & $\B_3$ \\
$Q^2$ &   &   &   &   &   &   &   \\
\hline 
\hline
quark  & 4.080 &  -0.673 & -0.440 & -0.707 & 0.196 & 1.717& 1.359          \\
2 GeV$^2$ &  &  &  &  &  &  &  \\
\hline
quark & 5.872 & -1.103 & -0.425 & -0.436 & 0.410 & 2.997  & 2.164 \\
109 GeV$^2$ &  &  &  &  &  &   &     \\
\hline
gluon  & 8.000 & -6.246 & -1.319 & 6.736 & 4.324 & 15.214 & 1.351    \\
2 GeV$^2$ &  &  &  &  &  &  &        \\
\hline
gluon & 1.090 & -8.848 & -1.430 & 8.568 & 6.216 & 22.031  & -0.145  \\
109 GeV$^2$ &  &  &  &  &  &  &   \\
\hline
\end{tabular}
\caption{ Values of different parameters used in the fit to the fragmentation functions.}
\label{tab1}
\end{table}

\begin{figure}[htb!]
  \resizebox{3.2in}{4.1in}{\includegraphics[0.5in,1in][7.5in,10in]{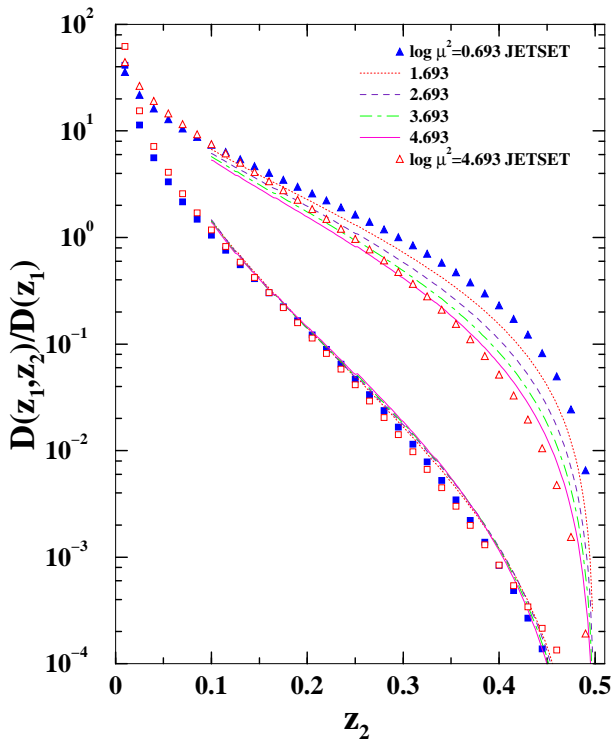}} 
    \caption{ Results of the evolution of the ratio of the dihadron fragmentation 
    function to the single fragmentation function for quarks and gluons.   }
    \label{fig10}
\end{figure}

In the recent experimental studies \eg, two high $p_T$ particle 
production in $p+p$, $p+A$ and $A+A$ collisions at RHIC, or 
two particle correlation in inclusive DIS by the HERMES experiment, 
one usually measures dihadron correlations in 
the form of inclusive spectra of associated particles produced 
in correlation with a high momentum trigger hadron. 
These measurements are essentially the ratios of the number of 
trigger and associated particle pairs divided by the number of triggers. 
They are equivalent to the ratios of the dihadron fragmentation 
functions to the single fragmentation functions. 
This quantity is plotted in Fig.~\ref{fig10} for the same range 
of scales as that in Fig~\ref{fig9}. The value of $z_1$ is once again 
held fixed at 0.5. The single hadron fragmentation functions
are given by BKK parameterization \cite{bin95} at same $Q^2$ 
which can also be derived from the same Monte Carlo simulations 
or evolved with the DGLAP equations for single hadron fragmentation functions. 
We have checked that all three methods give almost identical results. 
While it may come as no surprise that the evolution of this 
ratio is also predicted by the dihadron DGLAP evolution equations, 
it must be pointed out that the ratio of the dihadron and single
hadron fragmentation functions do not display substantial change 
with variation of the scale. This is especially true for the 
gluon fragmentation function.

\section{Discussions and Conclusions}

In this study, a very general discussion of the properties 
of the dihadron fragmentation functions has been carried out. 
Various results have been outlined: the momentum sum rules have 
been discussed, evolution equations for both the singlet quark 
and gluon dihadron fragmentation functions have been derived 
and solved numerically, comparisons with results from Monte Carlo event 
generators (in the absence of experimental measurements) are made.
Very good agreement has been obtained from such comparisons.
In the remaining, we cast a backward glance at dihadron fragmentation 
functions, summarizing the sentinel points of the aforementioned 
analysis, and present an overview of future studies.

Factorization ensures that fragmentation functions are well 
defined and universal. Once defined and measured in a 
given process they must admit the 
same definition and hence numerical value in another experiment.
Such definitions in light-cone gauge were presented in
Ref.~\cite{maj04a} in LO of QCD.
Gauge invariance of the definition is apparent
and the generalization involves little more than the 
similar procedure invoked in the case of the single inclusive 
fragmentation function \cite{col89}: The partonic operators at 
two space time points were connected by a Wilson line of the gauge field.
In the case of the gluon fragmentation function, the gluon vector 
potential was replaced with gauge invariant field operators.

Single inclusive fragmentation functions, which have the
interpretation of single inclusive hadron distributions, obey well
defined sum rules. The momentum sum rules for the dihadron 
fragmentation functions, which have the interpretation of the 
pair multiplicity in a jet, have been derived in Sec. III.
One can relate the dihadron to single hadron fragmentation functions
through these sum rules. However, they will involve first and second
order cumulant moments of the multiplicity distribution since 
dihadron fragmentation functions involve pair multiplicity.
The appearance of the cumulant moments makes the sum rules for
dihadron fragmentation functions less rigorous since they are
not well defined in the collinear factorization approximation.
However, in practice, they will provide useful guidance for 
phenomenological modeling of the dihadron fragmentation functions.

Though fragmentation functions are well defined in terms of parton
matrix elements, yet they still involve non-perturbative objects.
Therefore, the exact form of a fragmentation function may not 
be estimated purely from QCD. They, however, can be measured 
at a given energy scale for a whole set of continuous forward 
momentum fractions. Once such a measurement is made, one may 
derive the value at any other higher scale via the evolution 
equations in pQCD. Comparing these with 
experimental results at higher scales amounts to a test of 
the QCD evolution equations. In this paper, the lack of 
experimental results was circumvented by the use of the JETSET 
event generator which is tuned to fit a variety of experimental 
data on \epem annihilation. Note that event generators can 
reproduce not only the single inclusive measurements in \epem 
events, but also a multitude of many particle observables.
We used dihadron fragmentation functions from JETSET simulations 
at the scale of $Q_0 = 2$ GeV$^2$ as the initial condition for
the DGLAP evolution equations. We demonstrated that the
results at higher scales, {\it e.g.}, $Q^2 = 109$ GeV$^2$ 
from the numerical solution to the evolution equations
are in excellent agreement with that extracted from JETSET
simulations at the same scale, as shown 
in Figs.~\ref{fig9} and ~\ref{fig10}. This comparison provided 
a stringent test to the new evolution equations presented in this study.

\begin{figure}[htb!]
  \resizebox{3.2in}{4.1in}{\includegraphics[0.5in,1in][7.5in,10in]{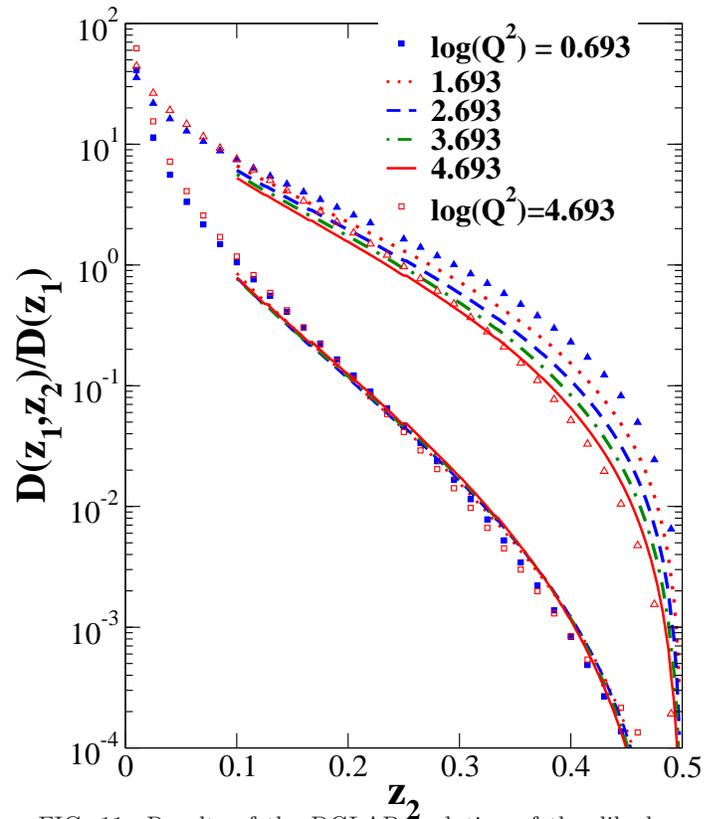}} 
    \caption{Results of the DGLAP evolution of the dihadron fragmentation 
    function without the independent fragmentation terms. All 
    parameters are the same as Fig.~\ref{fig9}.}
    \label{fig11}
\end{figure}

The effect and validity of each component of the DGLAP equations may 
be further tested by a simple exercise. The various terms presented in 
Eqs.~(\ref{q_dglap}) and (\ref{g_dglap}) may be broadly divided into two 
categories. The regular components, which we call {\it correlated} 
two hadron fragmentation, are contained in the first 
and last lines of Eq.~(\ref{q_dglap}) and the first and fourth lines
of Eq.~(\ref{g_dglap}). These are simple generalizations of
the single inclusive DGLAP equations and depend solely on dihadron
fragmentation functions. The new components, which we call
{\it independent} two hadron fragmentation, make up the second
and third line of  Eq.~(\ref{q_dglap}) and the second, third and last 
line of Eq.~(\ref{g_dglap}). These depend on products of 
single inclusive fragmentation functions, and therefore couple 
the evolution of the dihadron fragmentation functions with that 
of the single fragmentation functions. We may compute the evolution 
of dihadron fragmentation functions without the contribution 
from the components of independent fragmentation. This
leads to the results presented in Fig.~\ref{fig11}. As the reader may
note, the calculated evolution is now a poorer fit to 
the results from JETSET. While this may not be very noticeable 
for the quark fragmentation function, the gluon fragmentation 
function which is at least an order 
of magnitude smaller than the quark fragmentation function shows a
very noticeable difference. The evolution of the gluon fragmentation 
function depends on two separate pieces of independent fragmentation. 
One receives contributions from the product of two gluon single fragmentation 
functions, while another from the product of a quark and an antiquark 
distribution functions. It is the removal of latter that produces the majority 
of the difference in the evolution of the gluon dihadron fragmentation
function between Fig.~\ref{fig9} and Fig.~\ref{fig11}. This is due to
the fact that we restrict our attention to large $z$ where the quark 
fragmentation function dominates over the gluon, and the product of 
two single quark fragmentation functions is larger than the gluon dihadron 
fragmentation function at large $z_1$ and $z_2$. Thus each component of
the evolution has a role to play in the change of the fragmentation 
function from one scale to another. The contributions from the
new components of independent fragmentation are always positive.
Therefore, they slow down the scale evolution of the dihadron
fragmentation functions, particularly for gluon jets at large $z_1$ and
$z_2$. This is also the reason why the triggered distributions (the
ratio between dihadron and single hadron fragmentation functions)
change very little with the scale for gluon jets as shown 
in Fig.~\ref{fig9}. 

With the question of evolution of the dihadron fragmentation functions 
set aside, we now focus on certain general properties of the fragmentation 
function as has been exposed by this analysis.
It may be noted from Fig.~\ref{fig10} that the ratio of the dihadron 
fragmentation function to the single
fragmentation function of the leading hadron  
($D_q^{h_1h_2}(z_1,z_2,Q^2) / D_q^{h_1}(z_1,Q^2) $ ) shows little change 
as a function of $Q^2$ even as $Q^2 \ra 100$ GeV$^2$. 
This is especially true of the ratio in the case of the gluon 
fragmentation function, which shows practically no change 
with $Q^2$ over the range of $z_1$ and $z_2$ explored. 
The results of evolution are strongly dependent, however, 
on the initial conditions and thus on the actual values of $z_1$ and 
$z_2$. This is consistent with the observations noted in our
previous study of quark non-singlet dihadron fragmentation 
functions \cite{maj04a}.

In recent experiments at RHIC \cite{adl03}, correlations of two high
$p_\perp$ hadrons have been measured in $p+p$, $d+Au$ and $Au+Au$
collisions. Hadrons with $p_\perp \geq 4 $ GeV are used as a trigger. 
Once such a ``leading'' particle is identified, the experiment 
measures the differential probability of the ``next-to-leading'' 
hadrons or associated high $p_\perp$ particles emanating 
from the same collision with transverse momentum in the 
range $2 < p_\perp < 4 $ GeV at a given azimuthal angle $\phi$ 
with respect to the direction of the leading particle. 
In Ref.~\cite{adl03}, results for $d N/d \phi/N_{trig}$ are 
measured. While a large suppression was noted for $\phi \approx \pi$ 
(away-side) in central collisions, almost no change was seen in the 
vicinity of $\phi \approx 0$ (near side). Assuming both hadrons 
(leading and associated) with the same-side correlation come from
fragmentation of a jet, the initial jet energy is at least
about 7 to 10 GeV. This is about the same range of momentum scale
we have considered in studying the evolution of dihadron fragmentation
functions. This range of $Q^2$ is also relevant to the experiments 
of DIS on nuclei \cite{din04}, where medium modification of the 
ratio of the dihadron fragmentation function to the single 
fragmentation function of the leading hadron was reported. 
A double ratio of the above quantity $R_A/R_D$ with a large 
nucleus versus that in deuteron showed minimal change with the 
atomic number. A complete understanding of the relevance of this
observation requires a repeat of the calculation presented 
in this article with the inclusion of a medium modification.
While the study with medium modification will be presented 
in a forthcoming article, we can already see the trend by 
analyzing the analogous effects of QCD evolution in the vacuum.

If the results of Ref. \cite{adl03}  for $dN/d\phi/N_{trig}$
are integrated over the angle $\phi$ in the vicinity of $\phi = 0$, 
one should obtain, in effect, the ratio of the dihadron fragmentation 
function to the single fragmentation function. 
This is, no doubt, based on the assumption that both the leading and 
the next-to-leading particles emerge from the same parent parton.
In the results of Ref. \cite{adl03} the ratio is essentially 
integrated from $z_2\approx 0.3$, $z_1\approx 0.6$ to
$z_2\approx 0.4$, $z_1\approx 0.4$. Within this kinematic range,
there is not much change of the ratio of the fragmentation functions 
as a function of the $Q^2$ due to gluon bremsstrahlung in vacuum.
Therefore, it may not be entirely surprising that no variation 
was noted in the same side two hadron correlations in Ref. \cite{adl03}
due to medium induced gluon bremsstrahlung, in particular
if one takes into account the trigger bias caused by parton
energy loss \cite{Wang:2003aw}.

\begin{figure}[htb!]
  \resizebox{3.2in}{4.1in}{\includegraphics[0.5in,1in][7.5in,10in]{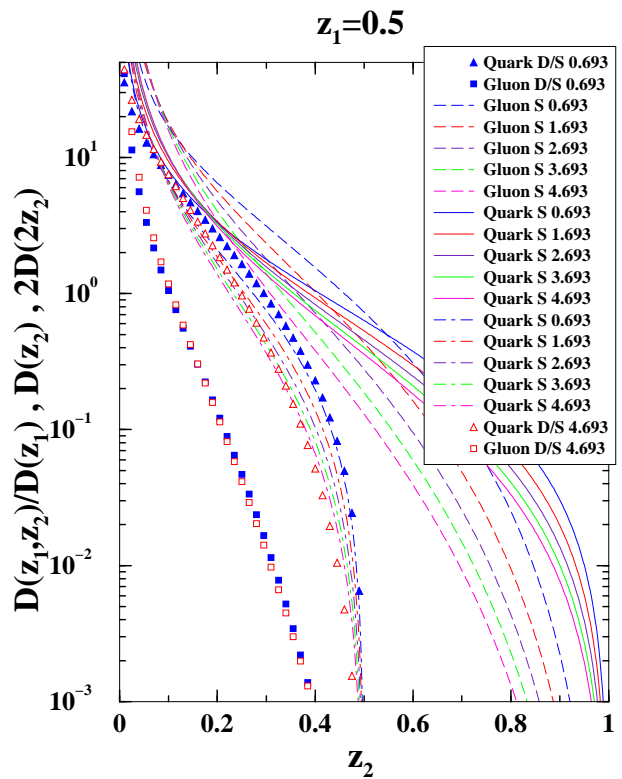}} 
    \caption{ Comparisons between the ratio of the dihadron fragmentation function to the 
    single fragmentation functions
    of the leading hadron (indicated as D/S) and the 
    fragmentation function of the associated hadron (indicated as S).
    The D/S curves are the same as in Fig.~\ref{fig10}. The single fragmentation 
    functions are obtained from the BKK parameterization of Ref.~\cite{bin95}. The 
    dashed lines are for the gluon and the solid lines for the quark fragmentation function.
    The dot-dashed lines represent a rescaled quark fragmentation function 
    $2D_q^{h_2}(2 z_2)$. The $Q^2$ of a particular dot-dashed line is the same 
    as the $Q^2$ of the solid line in the same order. }
    \label{fig12}
\end{figure}

It has been pointed out recently that the ratio of the dihadron 
fragmentation function to the single fragmentation function of the 
leading hadron (\tie, $D(z_1,z_2)/D(z_1)$) may be numerically 
similar to the single fragmentation function of the associated 
hadron \tie, $D(z_2)$ \cite{jia04}. This was noted experimentally and 
in simulations. We point out that there is indeed some truth to 
this observation. In Fig.~\ref{fig12} we plot the ratios of the 
dihadron fragmentation function to the fragmentation function of 
the leading hadron for the quark and gluon. This is compared 
with the single inclusive fragmentation functions of a 
quark (solid lines) and gluon (dashed lines).
In the range $0<z_2<0.3$, the single inclusive fragmentation functions
do indeed closely approximate the ratio $D_q(z_1,z_2)/D_q(z_1)$. 
This fact is however only true for the quark; no such similarity 
is noted for the gluon. The primary reason for the difference 
between $D_q(z_2)$ and $D_q(z_1,z_2)/D_q(z_1)$ 
for $z_2 > 0.3$ is the differences in kinematic bounds 
experienced by the two quantities: $D_q(z_2) \ra 0$ as $z_2 \ra 1$ 
whereas $D_q(z_1,z_2) \ra 0$ as $z_2 \ra 1 - z_1$. 
In Fig.~\ref{fig12}, $z_1$ is held fixed at 0.5, thus the 
maximum value of $z_2=0.5$. This differences in kinematic 
bounds may be circumvented by considering a rescaled single 
fragmentation function $D_q(z_2/(1-z_1))/(1-z_1)=2D_q(2z_2)$ 
which experiences a similar kinematic bound 
as  $D_q(z_1,z_2)/D_q(z_1)$ for $z_1=0.5$. This rescaled 
function is represented by the dot-dashed lines in Fig.~\ref{fig12}
which is very similar to $D_q(z_1,z_2)/D_q(z_1)$ with
the same momentum scale $Q$.

Finally, we point out that in the derivation of results presented 
in this paper, we depended on the assumption that 
the energy scales of the processes in question were high enough for the 
applicability of pQCD methods. We also required the fragmentation 
functions to be defined at a scale larger than a semihard 
scale $\mu_\perp$, such that $\Lambda^2_{QCD} << \mu^2_\perp << Q^2$. 
This semi-hard scale $\mu_\perp$ restricts the relative
transverse momentum between hadrons such that they are
considered as from one parton jet.
We thus required  $Q^2$ to be high enough for a hierarchy of 
scales to exist. The evolution equations are applicable solely 
when this hierarchy of scales exists.

\begin{acknowledgments}
The authors wish to thank E. Wang for many helpful discussions. 
This work was supported 
in part by the Natural Sciences and 
Engineering Research
Council of Canada, 
and in part 
by the Director, Office of Science, Office of High Energy and Nuclear Physics, 
Division of Nuclear Physics, and by the Office of Basic Energy
Sciences, Division of Nuclear Sciences, of the U.S. Department of Energy 
under Contract No. DE-AC03-76SF00098. 
\end{acknowledgments}

\end{document}